\documentclass[preprint, showpacs, preprintnumbers, amsmath,amssymb]{revtex4}

\usepackage{graphicx}
\usepackage{dcolumn}
\usepackage{tensor}
\usepackage{fontenc}

\begin{document}

\title{ The Friedmann equation in modified entropy-area relation from entropy force }
\author{Bin Liu}
\author{Yun-Chuan Dai}
\author{Xian-Ru Hu}
\author{Jian-Bo Deng}
\email{dengjb@lzu.edu.cn} \affiliation{
Institute of Theoretical Physics, Lanzhou University \\
Lanzhou 730000, People's Republic of China}

\date{\today}

%\keywords{HL gravity, Friedmann equation, Debye model.}
\begin{abstract}

According to the formal holographic principle, a modification to the assumption of holographic principle in Verlinder's
investigation of entropy force is obtained. A more precise relation between entropy and area in the holographic system is proposed.
With the entropy corrections to the area-relation, we derivate Newton's laws and Einstein equation
with a static spherically symmetric holographic screen. Furthermore we derived the correction terms to the
modified Friedmann equation of the FRW universe starting from the holographic principle and the Debye model.

%\pacs{04.70.Dy}
\pacs{04.70.Dy, 45.20.Dd}
\textbf{Keywords}:{ holographic principle, Friedmann equation, Debye model.}

\end{abstract}
\maketitle

\section{introduction}

The law of gravity, which was discovered by Isaac Newton in the 18th century and reformulated by Albert Einstein in the early 20th century,
still remains unclear in the microscopic level nowadays. Enlightened by Hawking et al.'s work \cite{f1}\cite{f2}\cite{f40}\cite{f3}
about the black hole entropy in 1970's,
Jacobson \cite{f4} got the Einstein's equations starting from the first law of thermodynamics. Similar research of Padmanabhan \cite{f5}\cite{f6}
also proposes gravity a thermodynamical interpretation. These results prompt people to take a statistical physics point of view on gravity. \\
\indent Recently, Verlinde \cite{f7} reinterpreted gravity as an entropic force caused by a change of amount of information associated
with the positions of bodies of matter. Lots of work has been done to reveal the implications of the entropic force interpretation of gravity.
The Friedmann equations and the modified Friedmann equations for Friedmann-Robertson-Walker universe
in Einstein gravity \cite{f8}\cite{f9}, f(R) gravity \cite{f10}, deformed Ho\v{r}ava-Lifshitz gravity \cite{f11} and braneworld scenario \cite{f12}
were derived
utilizing the holographic principle and the equipartition rule of energy. The Newtonian gravity in loop quantum gravity was given
by entropic force in Ref. \cite{f13}. In Ref. \cite{f14} the Coulomb force was regarded as an entropic force.
Ref. \cite{f15} showed that the holographic dark energy can be derived from the entropic force formula.
J. Makela pointed out in \cite{f16} that Verlinde's entropic force is actually the consequence of a specific microscopic model of spacetime.
The similar ideas were also applied to the construction of holographic actions from black hole entropy \cite{f17} while Ref. \cite{f18} showed that
gravity has a quantum informational origin. In \cite{f19}, a modified entropic force in the Debye model was presented.
Another important discovery was made by Joakim Munkhammar more recently \cite{f20},
which indicates a deep connection between the quantum mechanical probability density distribution and entropy.
For other relevant works to entropic force, we refer to e.g. \cite{f21}\cite{f42}\cite{f43}\cite{f44}\cite{f45}\cite{f46}\cite{f47}\cite{f48}
\cite{f22}\cite{f49}\cite{f50}\cite{f23}\cite{f51}\cite{f52}\cite{f53}\cite{f24}\cite{f25}\cite{f26} and references therein.\\
\indent According to Verlinde's discussion, the number of bits on holographic screen can be specified as $N=\frac{Ac^3}{G\hbar}$.
However, the entropy-area relation can be modified from the inclusion of quantum effects.
In this paper, by employing this modified entropy-area relation, we derive corrections to Newton's law of gravitation,
Einstein's equations, as well as the modified Friedman equations by adopting the viewpoint that gravity can be emerged as an entropic force.
The paper is organized as follows. Firstly, with a brief review of the idea of holographic principle and entropy,
a few entropy-area relations are discussed and a new relation is presented.
Second, we derivated Newton's law of gravity and Einstein equation followed the logic of Verlinde's derivation.
Third, motivated by modified entropy force model,
another modified Friedmann equation is proposed with the consideration of the new entropy-area relation.
Conclusions and further discussions can be found in part V. Here we set the units of $c=\hbar=k_B=1$.

%Sec.\ref{SecB}, we review the equations of the motion of a planet.
%In Sec.\ref{SecC}, choosing a Lagrangian function, we deduce the

\section{THE ENTROPY AND THE NUMBER OF BITS ON HOLOGRAPHIC SCREEN}\label{SecB}

We first review the idea of the entropy and the number of bits on holographic screen.
Assuming that gravity is sufficiently weak to allow for a choice of time slicing such that the matter system is at rest and
space is almost Euclidean. For any weakly gravitating matter system in asymptotically flat at space,
Bekenstein \cite{f27} has argued that the GSL (generalized second law of thermodynamics) implies the following bound
\begin{equation}\label{eq1}
S_{matter}\leqslant2\pi ER
\end{equation}
Here, $E$ is the total mass-energy of the matter system. The circumferential radius $R$ is the radius of the smallest sphere that fits around the matter system.\\
\indent Susskind \cite{f28} has argued that the GSL, applied to this transformation, yields the spherical entropy bound:
\begin{equation}\label{eq2}
S_{matter}\leqslant\frac{A}{4l^2_p}
\end{equation}
Here $A$ is a suitably defined area enclosing the matter system.\\
\indent Motivated by Bekenstein's entropy bound, recently Verlinde
postulated that on the holographic surface there live a set of
``bytes" of information which is called the number of degrees of
freedom and proportional to the area of the surface \cite{f7}
\begin{equation}\label{eq3}
N=S=\frac{A}{l^2_p}=\frac{Ac^3}{G\hbar}=\frac{A}{G}
\end{equation}
\indent With this in mind, some applications have been made followed the logic of Verlinde's derivation.
First in Ref. \cite{f29} two well-motivated corrections to the area-entropy relation have been considered,
which are the log correction and the volume correction
\begin{equation}\label{eq4}
S=\frac{A}{4G}-alog\left(\frac{A}{G}\right)+b\left(\frac{A}{G}\right)^\frac{3}{2}
\end{equation}
They shown that the log correction appears to have universality
emerged from the counting of microstates in many independent quantum
gravity theories and the ``volume" correction is also motivated by a
model for
the microscopic degrees comprising the black hole entropy in Loop Quantum Gravity.\\
\indent While considering the non-gravitational collapse condition, entropy-area relation would be another form \cite{f30}:
\begin{equation}\label{eq5}
S\leqslant\left(\frac{A}{G}\right)^\frac{3}{4}
\end{equation}
One could find this entropy bound, they indicated, if excluding configuration whose energies are so large
that they inevitably undergo gravitational collapse. Information entropy requires the energy,
while formation of a horizon by gravitational collapse restricts the amount of energy allowed in a finite region.\\
\indent  However symmetry is crucial aspect of any quantum approach. In \cite{f31} they have considered the quantum
effect in counting the degree of freedom on the screen. Assume that bits are distributed randomly on the lattice and
typically the variables could be with elementary spin $1/2$ of $SU(2)$ group, the number of degree of freedom
on the holographic screen becomes:
\begin{equation}\label{eq6}
N=S=\frac{A}{G}\left[1-\frac{3}{2}\left(ln\frac{A}{4G}\right)/\left(\frac{A}{4G}\right)\right]
\end{equation}
\indent Many other authors also suggest $S = A/4$ for some general holographic screen from different aspects \cite{f32} \cite{f41}.
Furthermore, recalling that there are various definitions of entropy-area relation,
it is time to put together all the pieces of thinking and see whether they can fit into a whole picture.
In this note we introduce a new parameter into Verlinde's work in order to correspond with the formal
holographic principle and we think entropy is information entropy.\\
\indent According to the above discussion, the entropy of a black hole is given by the Bekenstein-Hawking formula $S=A/4$.
In this case, $A/4$ as an area is more meaningful than $A$.
Because the entropy $S$ depends on the gravity theory and takes different forms for different gravity theories,
it is reasonable to discuss that there exists a logarithmic term in the entropy/area relation $S=\frac{A}{4}+\beta \ln A$
for the deformed Ho\v{r}ava-Lifshitz (HL) gravity \cite{f37}\cite{f38}\cite{f39}.
Motivated by deformed HL gravity, the corrected entropy is of the form:
\begin{equation}\label{eq7}
S=N=\frac{A}{4G}+\beta \ln\left(\frac{A}{G}\right)
\end{equation}
Here $\beta$ is dimensionless constant of order unity whose value is currently to be determined in Ho\v{r}ava-Lifshitz gravity.
In Eq.(\ref{eq2}) if let $V$ be a compact portion of a hyper surface of equal time in the spacetime $M$ and let $S(V)$
be the entropy of all matter systems in $V$. $B$ is the boundary of $V$ and $A$ is the area of the boundary of $V$. Then:
\begin{equation}\label{eq8}
S_V\leqslant \frac{A[B(V)]}{4l^2_p}
\end{equation}
\indent The entropy contained in any spatial region will not exceed the area of the region's boundary,
which is called spacelike entropy bound. All information about a gravitational system in a spatial region is encoded in its boundary \cite{f33},
so $S_V$ is the information entropy.\\
\indent It is known that the number of degrees of freedom $n$:
\begin{equation}\label{eq9}
n=ln\mathcal{N}=ln [dim(\mathcal{H})]
\end{equation}
For a quantum-mechanical system, $n$ is the logarithm of the dimension $\mathcal{N}$ of its Hilbert space $\mathcal{H}$.
Now consider a $3$ dimensional lattice of spin like degrees of freedom.
For definiteness assume the lattice spacing is the Planck length and that each site is equipped with a spin which can be in one of two states.
Think over $N_{bit}$ is the number of sites in a region of space of volume $V$ of distinct orthogonal quantum states, one can have:
\begin{equation}\label{eq10}
\mathcal{N}(V)=2^{N_{bit}}
\end{equation}
Note that $N_{bit}$ is defined as the number of bits, which is proportional to the thermodynamic entropy.
The logarithm of $\mathcal{N}(V)$ is the number of degrees of freedom in $V$ and satisfies:
\begin{equation}\label{eq11}
n=ln\mathcal{N}=N_{bit}(ln2)
\end{equation}
Here $n$ is proportional to the information entropy regarding the system is encoded.
It is up to a factor of $(ln2)$ to the number of bits of information needed to characterize a state.
The hypothesis that gravity is fundamentally thermodynamic is essentially that we take this entropy seriously as thermodynamic entropy,
which is as a measure of disorder. We consider a situation where $S$ can be regarded as a two-sphere of a given radius $R$,
and we assume that there is a spherically symmetric mass distribution in the interior region,
which is approximately static and in equilibrium. As above example, if every Planck area has possibilities of $f$,
the number of bits would become:
\begin{equation}\label{eq12}
N_{bit}=\frac{n}{(lnf)}
\end{equation}
Here the number of bits $N_{bit}$ would be understood as the number of Planck area on holographic surface.
Considering entropy as thermodynamic entropy, then there is no logarithmic term, and $N$
(the number of degrees of freedom) in Eq.(\ref{eq7}) is equivalent to $n$ in Eq.(\ref{eq9}).
While considering entropy is information entropy, in order to make our discussion follow the thermodynamic method,
$N$ in Eq.(\ref{eq7}) is equivalent to $n$ in Eq.(\ref{eq11}) and should be substituted with this new definition $N_{bit}$
\begin{equation}\label{eq13}
N_{bit}=\frac{A}{4(lnf)G}+\beta ln\left[\frac{A}{(lnf)G}\right]
\end{equation}
This is different from the relation given by Verlinde in \cite{f7}. Logarithmic terms and the
coefficient of $1/4$ are added to the relation which follows from the clearness of the assumptions.
In the following, we will derivate Newton's law of gravitation and Einstein equation with a static spherically
symmetric holographic screen according to the expression (\ref{eq13}).\\

\section{DERIVATION OF NEWTON'S GRAVITATIONAL LAW AND EINSTEIN EQUATION}\label{SecC}

\indent It would be interesting to examine how the number of bits on the holographic screen will influence the equation of gravity.
We now have the ingredients necessary to run a version of Verlinde's argument. \\
\indent Considering a particle with mass $m$ approaching a small piece of holographic screen from the side,
Verlinde concluded that the entropy associated with this process should be Bekenstein entropy \cite{f7}
\begin{equation}\label{eq14}
\Delta S=2\pi m \Delta x
\end{equation}
\indent Then according to the equipartition law of energy, the total energy on the surface is:
\begin{equation}\label{eq15}
E=M=\frac{1}{2}N_{bit}T
\end{equation}
\indent From Eq.(\ref{eq13}), it is easy to arrive at:
\begin{equation}\label{eq16}
N_{bit}=\frac{A}{4(lnf)G}\left[1+\beta \left(ln\frac{A}{(lnf)G}\right)/\left(\frac{A}{4(lnf)G}\right)\right]
\end{equation}
Comparing (\ref{eq14}) (\ref{eq15}) and (\ref{eq16}),
we can obtain the entropic force due to the change of the virtual information on the screen:
\begin{equation}\label{eq17}
F=T\frac{\Delta S}{\Delta x}=-\frac{GMm}{R^2}\frac{4(lnf)}{1+\beta \left(ln\frac{A}{(lnf)G}\right)/\left(\frac{A}{4(lnf)G}\right)}
\end{equation}
So we clearly see that the number of bits brings the correction to the entropic force.
There are logarithmic terms which mean that they could also cause the change of Newtonian gravitational force.
Macroscopically, when we consider $A/G\gg1$ and $\left(ln\frac{A}{(lnf)G}\right)/\left(\frac{A}{4(lnf)G}\right)\ll1$,
this correction will be neglected. \\
\indent We can further consider the influence on the Einstein equation caused by the number of bits on the holographic screen.
Differentiate Eq.(\ref{eq13}):
\begin{equation}\label{eq18}
d N_{bit}=\left(\frac{1}{4(lnf)G}+\frac{\beta}{A}\right)dA
\end{equation}
Expressing the energy in terms of the total enclosed mass $M$ and employing the law of equipartition, we have:
\begin{equation}\label{eq19}
M=\frac{1}{2}\int_STdN_{bit}=\frac{1}{4\pi G}\int_S\left(\frac{1}{4(lnf)}+\frac{\beta G}{A}\right) e^{\phi}\nabla{\phi}d A
\end{equation}
Following the same logic in \cite{f7}, we can get the integral relation
\begin{equation}\label{eq20}
2\int_{\Sigma}(T_{\mu \nu}-\frac{1}{2}Tg_{\mu \nu}) n^\mu \xi^\nu d V=\frac{1}{16\pi G(lnf)}\int_{\Sigma}R_{\mu \nu}n^\mu \xi^\nu d V+\frac{\beta}{4\pi}\int_S\frac{e^{\phi}\nabla{\phi}}{A}dA
\end{equation}
The second term on the right-hand-side is an additional term compared with Verlinde's result in \cite{f7}.
This term is caused by the correction to the number of bits on the holographic screen which brings a surface correction to the Einstein equation. For the Schwarzschild spacetime:
\begin{equation}\label{eq21}
\int_S\frac{e^{\phi}\nabla{\phi}}{A}dA=4\pi \int_S \frac{GM}{4\pi R^2} \frac{dA}{A}=-\frac{4\pi MG}{A}
\end{equation}
And the Eq.(\ref{eq20}) becomes:
\begin{equation}\label{eq22}
2\int_{\Sigma}(T_{\mu \nu}-\frac{1}{2}Tg_{\mu \nu}) n^\mu \xi^\nu d V=\frac{1}{16\pi G(lnf)}\int_{\Sigma}R_{\mu \nu}n^\mu \xi^\nu d V-2\int_{\Sigma}\frac{\beta G}{A}(T_{\mu \nu}-\frac{1}{2}Tg_{\mu \nu}) n^\mu \xi^\nu d V
\end{equation}
Here we use $M=2\int_{\Sigma}(T_{\mu \nu}-\frac{1}{2}Tg_{\mu \nu}) n^\mu \xi^\nu d V$ and $\frac{\beta G}{A}\ll1$ in our discussion.
We derive corrections to the Einstein equation:
\begin{equation}\label{eq23}
R_{\mu \nu}=32\pi G(lnf)(T_{\mu \nu}-\frac{1}{2}Tg_{\mu \nu})+M_{\mu \nu}
\end{equation}
where $M_{\mu \nu}=\frac{32\pi \beta G^2(lnf)}{A}(T_{\mu \nu}-\frac{1}{2}Tg_{\mu \nu})$
is the correction which is a part of matter field causing curved spacetime. When we consider $A/G\gg1$,
$M_{\mu \nu}$ become so small that this correction could be neglected.
Different from the usual field equations, the term $M_{\mu \nu}$ , which is from the surface term,
is a nonlocal effect determined by the holographic description of boundary physics in the frame of holography.
The obtained additional surface term in the Einstein equation brings the similarity to the surface term in the Einstein
equation discussed in \cite{f34}\cite{f35}.

\section{DERIVATION OF FRIEDMANN EQUATION}\label{SecD}

In this section, by employing above modified entropy-area relation (\ref{eq13}),
we reproduce the modified Friedmann equation motivated by methods in \cite{f26} and we adopt
the viewpoint that gravity can be emerged as an entropic force.
The homogeneous and isotropic universe model is described by the metric:
\begin{equation}\label{eq24}
ds^2=-dt^2+a^2(t) \left (\frac{dr^2}{1-kr^2}+r^2d\Omega^2 \right )
\end{equation}
here $d\Omega^2$ is the line element of a two-dimensional unit sphere. Adopting a new coordinate $R=a(t)r$,
$a$ is the scale factor, the metric (\ref{eq24}) will become:
\begin{equation}\label{eq25}
ds^2=h_{ab}dx^adx^b+R^2d\Omega^2
\end{equation}
With $x^0=t,x^1=r,h_{ab}=diag(-1,a^2/(1-kr^2))$ , the dynamical apparent horizon $R_A$ is determined by
$h^{ab}\partial_aR\partial_bR=0$, and is given by:
\begin{equation}\label{eq26}
R_A=ar=\frac{1}{\sqrt{H^2+k/a^2}}
\end{equation}
here $H=\dot{a} / a$ is the Hubble parameter. Suppose that the energy-momentum tensor $T_{\mu \nu}$
of the matter in the universe has the form of a perfect fluid $T_{\mu\nu}=(\rho+p)u_\mu u_\nu+pg_{\mu \nu}$,
where $u^{\mu}$ denotes the four-velocity of the fluid and $\rho$ and $p$ are the energy density and pressure, respectively.
The energy conservation law $\nabla_\mu T^{\mu \nu}=0$ gives the continuity equation in the form:
\begin{equation}\label{eq27}
\dot{\rho}+3H(\rho+p)=0
\end{equation}
Consider a compact spatial region $V$ with a compact boundary $S$, which is a sphere with physical radius
$R=a(t)r$. The active gravitational mass is defined as \cite{f9}:
\begin{equation}\label{eq28}
M=2\int_v(T_{\mu\nu}- \frac{1}{2} Tg_{\mu\nu})u^\mu u^\nu\mathrm{d}V=\frac{4\pi a^3 r^3}{3} (\rho+3p)
\end{equation}
\indent In section II, the corrected entropy we derivated takes the form:
\begin{equation}\label{eq29}
N_{bit}=S=\frac{A}{4(lnf)G}+\beta ln\left[\frac{A}{(lnf)G}\right]
\end{equation}
To derive the entropic law, the surface $S$ is between the test mass and the source mass,
but the test mass is assumed to be very close to the surface as compared to its reduced Compton wavelength.
When a test mass $m$ is a distance $\Delta x=\lambda_m$ away from the surface $S$,
the entropy of the surface changes by one fundamental unit $\Delta S$ fixed by the discrete spectrum
of the area of the surface via the relation:
\begin{equation}\label{eq30}
dS=\frac{\partial S}{\partial A}dA=\left(\frac{1}{4(lnf)G}+\frac{\beta}{A}\right)dA
\end{equation}
Because ``bits" of information is scale proportional to the area of
the surface, it is appropriate to get:
\begin{equation}\label{eq31}
A=QN_{bit}
\end{equation}
Where $Q$ is a fundamental constant which should be specified later.
Note that $N_{bit}$ is the number of bits and thus $dN_{bit}=1$, we have $dA=Q$. Hence from (\ref{eq30}), we reach:
\begin{equation}\label{eq32}
dS=\left(\frac{1}{4(lnf)G}+\frac{\beta}{A}\right)Q
\end{equation}
\indent Before employing the equipartition law of energy, it should be noticed that it does not hold at very low temperatures,
and the equipartition law of energy derived from Debye model is in good agreement with experimental results.
Since the equipartition law of energy plays an important role in the derivation of entropic force,
it should be modified for the very weak gravitational fields which corresponds to very low temperature \cite{f19}
\begin{equation}\label{eq33}
E=M=\frac{1}{2}\ N_{bit}TD(x)
\end{equation}
Where the one dimensional Debye function is defined as:
\begin{equation}\label{eq34}
D(x)=\frac{3}{x^3}\int^x_0 \frac{y^3}{e^y-1}dy
\end{equation}
And $x$ is related to the temperature $T$ with Unruh temperature \cite{f36}
\begin{equation}\label{eq35}
x=\frac{T_D}{T}=\frac{g_D}{g}
\end{equation}
It shows that when in strong gravitational fields, we have $x\ll1$ and $D(x)\rightarrow1$.
In other words, it implies that the effect of Debye model would be very weak in the early universe.
In \cite{f19}, Gao used the three dimensional Debye model to modify the entropic force.
The MEF can be very well approximated by the Newtonian gravity in the solar system and in the Galaxy.
Especially, the large-scale universe is in such an extreme weak gravitational field, and hence the MEF makes sense in cosmology.
With this Debye acceleration, the current cosmic speeding up can be explained without invoking any kind of dark energy.
Other studies following MEF model, we refer to \cite{f21} for an incomplete list.
So the temperature on the surface is:
\begin{equation}\label{eq36}
T=\frac{MQ}{2\pi R^2 D(x)}
\end{equation}
Now, we are in a position to derive the entropic-corrected Newton's law of gravity by the
thermodynamic equation of state. Combining Eqs.(\ref{eq32}) and (\ref{eq36}), we easily obtain:
\begin{equation}\label{eq37}
F=T\frac{\partial S}{\partial x}=-\frac{Mm}{R^2}\frac{Q^2}{2\pi D(x)}\left(\frac{1}{4(lnf)G}+\frac{\beta}{A}\right)
\end{equation}
This is nothing but the Newton's law of gravitation to the first order if we define $Q^2=8\pi (lnf)G^2$.
Thus we obtain the modified Newton's law of gravitation as:
\begin{equation}\label{eq38}
F=-\frac{GMm}{R^2}\frac{1}{D(x)}\left(1+\frac{\beta(lnf)G}{\pi R^2}\right)
\end{equation}
Thus we see that the Newton's law of gravity will modified accordingly with the correction
in the entropy-are expression. As we mentioned, these corrections are well motivated from gravity theories.\\
\indent On the other hand, dynamical equation for Newtonian cosmology should be first derived.
Note that here $r$ is a dimensionless quantity which remains constant for any cosmological object
partaking in free cosmic expansion. Combining the second law of Newton for the test particle $m$ near the surface,
we use gravitational force (\ref{eq38}) then get:
\begin{equation}\label{eq39}
F=m\ddot{R}=m\ddot{a}r=-\frac{GMm}{R^2}\frac{1}{D(x)}\left(1+\frac{\beta(lnf)G}{\pi R^2}\right)
\end{equation}
Assume $\rho=M/V$ is the energy density of the matter inside the volume $V=\frac{4}{3}\pi R^3$.
Then, Eq.(\ref{eq39}) can be rewritten as:
\begin{equation}\label{eq40}
\frac{\ddot{a}}{a}D(x)=-\frac{4\pi G}{3}\rho\left(1+\frac{\beta(lnf)G}{\pi R^2}\right)
\end{equation}
This is nothing but the entropy-corrected dynamical equation for Newtonian cosmology.
The main difference between this equation and the standard dynamical equation for Newtonian cosmology is that
the correction terms depend on $D(x)$ and $R$. If we consider a flat universe (k=0), we get $R=1/H$ and Eq.(\ref{eq40})
could be rewritten in the form:
\begin{equation}\label{eq41}
\frac{\ddot{a}}{a}D(x)=-\frac{4\pi G}{3}\rho\left[1+\frac{\beta(lnf)G}{\pi}\left(\frac{\dot{a}}{a}\right)^2\right]
\end{equation}
Because $\rho=M/V$, we have:
\begin{equation}\label{eq42}
M=-\frac{\ddot{a}a^2r^3}{G}D(x)\left(1+\frac{\beta(lnf)G}{\pi R^2}\right)^{-1}
\end{equation}
Equating Eqs.(\ref{eq28}) and (\ref{eq42}) we find:
\begin{equation}\label{eq43}
\frac{\ddot{a}}{a}D(x)=-\frac{4\pi G}{3}(\rho+3p)\left(1+\frac{\beta(lnf)G}{\pi R^2}\right)
\end{equation}
This is the modified equation for the dynamical evolution of the FRW universe.
Multiplying $\dot{a}a$ on both sides of Eq.(\ref{eq43}), and using the continuity equation (\ref{eq27}),
we integrate the resulting equation and obtain:
\begin{equation}\label{eq44}
H^2=\left(\frac{\dot{a}}{a}\right)^2=\frac{8\pi G}{3}\rho \left[\frac{1}{D(x)}+\frac{\beta(lnf)G}{D(x)\pi R^2 \rho}\int \frac{d(\rho a^2)}{a^2}\right]
\end{equation}
In order to calculate the integrations in the correction terms we assume the equation of state parameter $\omega=p/\rho$ is a constant,
then the continuity equation can be integrated immediately to give:
\begin{equation}\label{eq45}
\rho=\rho_0a^{-3(1+\omega)}
\end{equation}
where $\rho_0$, an integration constant, is the present value of the energy density.
Inserting relation (\ref{eq45}) in Eq.(\ref{eq44}) and integrate, we obtain:
\begin{equation}\label{eq46}
H^2=\frac{8\pi G}{3}\rho \left[\frac{1}{D(x)}+\frac{\beta(lnf)G(1+3\omega)}{3D(x)\pi R^2(1+\omega) }\right]
\end{equation}
In this way we derive another modified Friedmann equation of FRW universe by considering
HL gravity and Debye model from entropic force.
It is clear that the number of bits and Debye model bring the correction to the Friedmann equation.
The additional term is caused but in absence of the correction terms ($\beta=0$),
one recovers the approximate MEF model \cite{f19}. When the universe becomes large ($a\gg1$),
Eq.(\ref{eq46}) would reduce to the approximate Gao's modified Friedmann equation.
Since the effect of this Debye model could be obvious enough,
it can sufficiently describe the accelerating universe without the assumption of dark energy.
However, the second term is comparable to first term when $a$ is very small,
the corrections make sense at early stage of the universe where $a\ll 1$.

\section{conclusion}\label{SecE}
\indent In summary, we have shown that with the entropy corrections to the area-relation of the black hole entropy,
the Newton's law of gravitation and the Einstein equation would be modified accordingly.
This correction is motivated from Ho\v{r}ava-Lifshitz theory and the information entropy has been considered.
Furthermore we derived the correction terms to the modified Friedmann equation of the FRW universe starting from
the holographic principle and the Debye model by using Verlinde's argument that gravity appears as an entropic force.
The Friedmann equation in the deformed entropy/area relation and Einstein gravity are very different in early universe,
but they give the same result in late universe. On the other hand, the modified Friedmann equation considering Debye model
and Einstein gravity vary in late universe, but agree with each other in early universe.
In particular, we have found that in early universe corrections from modified entropy-area relation
would be much stronger than the effect of MEF model. Contrarily, in late universe it would degenerate to MEF model
under certain conditions.
The new equation could be better in describing the whole evolution of the universe.
Our results may be useful to further investigations of the holographic properties and modified Friedmann equations for different gravity theories.
We stress the generality of these results and our study provides a strong consistency check on Verlinde's model.

\end{document}